# Restricted Isometry Random Variables: Probability Distributions, RIC Prediction and Phase Transition Analysis for Gaussian Encoders

Oliver James and Heung-No Lee*, *Senior Member, IEEE*

*Abstract*— In this paper, we aim to generalize the notion of restricted isometry constant (RIC) in compressive sensing (CS) to *restricted isometry random variable* (RIV). Associated with a deterministic encoder there are two RICs, namely, the left and the right RIC. We show that these RICs can be generalized to a left RIV and a right RIV for an ensemble of random encoders. We derive the probability and the cumulative distribution functions of these RIVs for the most widely used i.i.d. Gaussian encoders. We also derive the asymptotic distributions of the RIVs and show that the distribution of the left RIV converges (in distribution) to the Weibull distribution, whereas that of the right RIV converges to the Gumbel distribution. By adopting the RIV framework, we bring to forefront that the current practice of using eigenvalues for RIC prediction can be improved. We show on the one hand that the eigenvalue-based approaches tend to overestimate the RICs. On the other hand, the RIV-based analysis yields precise estimates of the RICs. We also demonstrate that this precise estimation aids to improve the previous RIC-based phase transition analysis in CS.

## I. INTRODUCTION

RESTRICTED isometry property (RIP) of an encoder [1] is a standard analysis tool in compressive sensing (CS). This property quantifies the goodness of an encoder for recovering sparse signals in terms of its restricted isometry constant (RIC). As a result, the RIC of an encoder is used to analyze the ability of a decoder for signal recovery. Thus, over the past eight years, finding the RIC of the encoders has received a great interest in the CS community [2][3].

We say an $N \times 1$ signal vector $x$ is $K$-sparse if it has *exactly* $K$ non-zero values. Let $\Psi$ denote the set of all $K$-sparse vectors and $\mathcal{S}$ represent the *support set* of $x$ which is a set of all indices that correspond to the non-zero values of $x$. Compressed measurements of $x$ are obtained by using an encoder $A$ as $y = Ax$, where $A$ is an $M \times N$ matrix with $M < N$. Sparse signal recovery problems deal with finding $x$ from only $M$ measurements, given the encoder $A$. Goodness of the encoder for signal recovery is studied using the RIP.

An encoder $A$ is said to obey the RIP with the smallest constants $0 < \delta_K^L < 1$ and $\delta_K^R > 0$ if the relation

$$1 - \delta_K^L \leq \frac{\|Ax\|_2^2}{\|x\|_2^2} \leq 1 + \delta_K^R \quad (1)$$

holds for every $x \in \Psi$. The parameter $\delta_K^L = 1 - \min_{x \in \Psi} \frac{\|Ax\|_2^2}{\|x\|_2^2}$ in (1) is called the left RIC and $\delta_K^R = \max_{x \in \Psi} \frac{\|Ax\|_2^2}{\|x\|_2^2} - 1$, the right RIC [4, p. 89] [5]. In [6], these RICs are referred to as the *asymmetric* RICs [6]. The *symmetric* constant $\delta_K = \max\{\delta_K^L, \delta_K^R\}$ with $0 < \delta_K < 1$ is referred to as the RIC [5, eq.(3)], which is easy to determine once the asymmetric RICs are known. Thus, in this paper, we deal with the general case of asymmetric RICs.

Verifying an encoder for an RIC is NP-hard and hence, it is computationally intractable [7][8]. Despite the computational aspect, RICs of encoders play a crucial role in finding the theoretical recovery guarantees for various decoders. For example, recently, Cai and Zhang [9] showed that if the symmetric RIC of an encoder is satisfies $\delta_K < \frac{1}{3}$, then the $L_1$ decoder exactly recovers every $K$-sparse signal. Such recent bound involving asymmetric RICs $\delta_K^L$ and $\delta_K^R$ is given in [3] (please see eq. 22). Several such RIC-based recovery conditions have been summarized in [2][10, p. 611] for various decoders. In addition, RICs are crucial for analyzing the stability of decoders under noise and for analyzing the phase transitions (PT) of the decoders [6][11]. Thus, various approaches [4][6][12-18] have been proposed to predict the behavior of the RICs (discussed in Sections I.2 and I. 3) and hence, recently, the area of RIC prediction has received a considerable attention.

*1. Major Contributions of this paper*

We aim to initiate, in this paper, a new direction of research by generalizing the RICs to the restricted isometry random variables (RIVs). This generalization opens up a brand-new approach to give unprecedented sharpness in RIC prediction. We list here the key contributions of our paper.

The authors are with the School of Information and Communications, Gwangju Institute of Science and Technology (GIST), South Korea (e-mail: heungno@gist.ac.kr; oliver@gist.ac.kr). This work was supported by the National Research Foundation of Korea grant funded by the Korean government (Do-Yak Research Program, No. 2013-035295).

1. *Generalization of RICs to RIVs (Section II):* Till date, for random decoders, the RICs are predicted via the probability distribution functions (PDFs) of the eigenvalue of the encoder [1][6]. In this paper, we introduce the concept of RIVs for RIC prediction. The RIVs generalize the RICs. For the Gaussian encoders, we define the RIVs precisely using the order statistics of the *Rayleigh quotient random variables* (see (5) and (6)). This course of thoughts enables us to derive the PDF of the RIVs directly using the concept of the order statistics of i.i.d. random variables.
2. *PDF of the RIVs for the Gaussian ensemble (Section III and IV):* We derive the PDF of the RIVs for the Gaussian encoders under two scenarios: Non-asymptotic and asymptotic. The convergence phenomena for these two scenarios are discussed in Section II. 3. For the non-asymptotic distributions, the PDFs are determined from the definitions of RIVs given in (5) and (6). The asymptotic PDFs are developed by borrowing the ideas from the *extreme value theory*. We show that the asymptotic PDF of the RIVs converges (in distribution) to either the Weibull or the Gumbel distribution. We are not aware of any previous work that introduces the RIVs or computes their PDFs via the extreme value theory. So, we provide a rudimentary treatment of the RIVs from the onset.
3. *RIC prediction and Phase Transition analysis (Section V):* On the application side of the new framework, we show the benefit of the RIVs in the analysis of two important problems in CS, namely, the RIC prediction and the phase transition (PT) analysis.

    RIC prediction is usually done by using the so-called *critical functions*. The current practice to derive the critical functions is by using the eigenvalue (EV) approach. Using the PDF of the RIVs, we derive new critical functions and show that the new functions are superior to the use of the existing EV-based functions (see Fig. 4). For detailed discussion on differences between the RIV and the EV approaches, please refer to Section V. 2.

    We use the new critical functions in order to derive the PT boundaries of $L_1$ decoders. We demonstrate that the proposed RIV-based approach leads to a remarkably improved PT boundary compared to the existing approaches (see Section V.4 for more details).

We next outline the history of RIC prediction and the related literatures that we use for comparison.

*2. Overview of Prior Work*

RIC prediction is the study of behavior of the RICs of an encoder. This study finds either 1) the level of measurements needed to obey the RIP with RIC less than one, or 2) the triplets $(K, M, N)$ for which the RICs are less than a predefined value.

**RIC prediction for deterministic encoders**: A deterministic encoder is the one in which the entries of the encoder are fixed, real or complex values. Two techniques for finding the RICs of the deterministic encoders are the Gershgorin circle theorem [18, p.1126] and the spark of an encoder [18, 1128]. The RICs of the encoders such as the cyclic encoders on finite fields [12], the chirp encoders [13], and the Fourier residual modulo encoders [14] are studied using the Gershgorin circle theorem. These encoders are shown to obey the RIP with the RIC less than one if the level of measurements is on the order of square of the sparsity, i.e., $M = \mathcal{O}(K^2)$.

The RICs of the encoders such as the Vandermonde [15] and the Steiner equiangular tight frames [16] are determined using their spark. Vandermonde encoders are shown to obey the RIP if all the bases are different. Under this condition, the level of measurements needed is $M = K+1$. However, when the encoder size is large, it is difficult to guarantee the RIP for these encoders. It is shown in [16] that for the Steiner equiangular tight frames, the level of measurement needed is also $\mathcal{O}(K^2)$.

However, for random encoders (which we discuss next) the measurement level needed is $\mathcal{O}(K \log^\alpha N), \alpha \geq 1$. This level is near-optimal as $M \geq 2K$ is the requirement for an encoder to obey the RIP with the RIC less than one. Thus, the deterministic encoders appear (theoretically) to be little inferior to that of the random encoders [17][18].

**RIC prediction for random encoders:** The entries of a random encoder are i.i.d. random variables. For random encoders, the RICs are predicted to have a good value in the asymptotic sense. For example, Candes and Tao [19] used the concentration inequalities to arrive at the RIP of random encoders such as the Gaussian [1, p 4209], the Bernoulli [19, p. 5414], and the Fourier with randomly sampled rows [19, p. 5415]. Rudelson and Vershynin used geometric approach to arrive at the RIP of the Fourier [20, Theorem 3.3] and the Gaussian encoders [20, Theorem 4.1]. Though the approaches in [19] and [20] are different, their conclusions are the same. That is, the measurement level needed for the Gaussian and the Bernoulli encoders is $\mathcal{O}(K \log N)$ and for the Fourier with randomly sampled rows it is $\mathcal{O}(K \log^6 N)$. The RIP of partial random circular encoders are reported in [21].

*3. Works related to this paper*

Apart from computing the required measurement level, in particular, for Gaussian encoders, various approaches [1][4][6][22] have been proposed to determine the triplets for which the encoder RIC is less than a prescribed value. In these methods, the central idea is to represent the RICs, in terms of the smallest and the largest eigenvalue of the matrix $A_{\mathcal{S}}^T A_{\mathcal{S}}$, where $A_{\mathcal{S}}$ denotes an $M \times K$ sub-matrix of $A$. That is, by supposing that the eigenvalues obey the relation,

$$(1-\delta_K^L) \leq \lambda_{\min}(A_{\mathcal{S}}^T A_{\mathcal{S}}) \leq \lambda_{\max}(A_{\mathcal{S}}^T A_{\mathcal{S}}) \leq (1+\delta_K^R) \quad (2)$$

a PDF is induced on the RICs via the PDF of the eigenvalues. The behavior of the RICs is then predicted by using the eigenvalue PDFs. We refer to this approach, in this paper, as the eigenvalue (EV) approach for the reason that a PDF is induced on the RICs via the PDF of the eigenvalues. The behavior of the RICs can be effectively visualized in the PT space.

**Definition 1** *Phase transition (PT) space:* Let $\rho := \frac{K}{M}$ be the signal sparsity measure and $\delta := \frac{M}{N}$ be the undersampling ratio. Then, the domain $(\rho, \delta) \in [0,1]^2$ is called the *PT space*.

**RICs prediction via critical functions:** The prediction of the left and the right RICs is usually studied, in the PT space, with the help of the *critical functions*. These functions provide an upper bound above which the RICs of the Gaussian encoders are unlikely to exist.

The earliest derivation of the critical functions is by Candes and Tao [1] using the concentration inequality on the PDF of the eigenvalues. Their critical function was substantially improved by Blanchard *et al.* [6] and Bah and Tanner [22] using an upper bound on the eigenvalue PDFs borrowed from the work of Edelman [23]. The critical functions in [6][22] are considered to be the best in terms of predicting the RICs (see [22, Fig. 2.2]). These functions do have an explicit closed-form, because they are determined from the solutions of algebraic equations [6, eqs. (2.5) and (2.6)].

In this paper, we aim to propose the RIV framework, which can offer much sharper RIC prediction results than the EV approach. For details please see Section V.3.

In [24], Stojnic proposed a different approach to estimate the RICs in order to circumvent the union-bounding strategy and the tail estimates of the EV approach. While the Stojnic approach I simpler than the EV, it gives similar estimate of the RICs as that of the EV as discussed in [2, Section 1, Section 2.2, Section 4.2]. Thus, it is good enough to compare our results with the EV approach, which we have done in Section V.

**RICs in the PT analysis:** PT is a systematic analysis framework for understanding the sparsity-undersampling tradeoffs of various encoder-decoder pairs [25]. It conveys practitioners a certain degree of assurance on how much undersampling can be made for a *K*-sparse signal while perfect recovery of the signal is not compromised. The central aim of PT analysis is to determine a precise PT boundary (in the PT space) that separates the successful recovery region from the failure recovery region. These boundaries have been given mostly empirically, but, there are analytical approaches, which will be discussed next.

The polytope approach is one such analytical study that was recently introduced by Dohono and Tanner [25]. They developed the PT boundaries for the Gaussian encoder and $L_1$ decoder pair by using the exponential bounds on the face counts of the polytopes [26][27]. Another geometric approach for the PT analysis is the geometric functional analysis (GFA) by Rudelson and Vershynin [20]. However, GFA is shown in [6] to be inferior to the polytope approach.

Apart from the above two geometric approaches, PT boundary can also be derived using the RIC prediction. This was first demonstrated by Blanchard *et al.* [6] using the EV approach. However, soon they discovered that EV-based approach could not precisely determine the PT boundary as predicted by the geometric approaches. The reason for such imprecision is unfortunately not known till date. Donoho and Tanner commented on this shortcoming of the EV approach in [28, Sections X and XI]. For comparison of PT boundaries determined by the polytope, the GFA, and the EV approaches, [6] and [28] can be referred. The EV-based PT analysis for greedy decoders are reported in [11]. The article [28] by Donoho and Tanner provides a good summary of recent approaches for finding the analytical PT boundaries. For an assorted number of compiled empirical and analytical PT boundaries for various encoder/decoder pairs, we refer the readers to [29-37].

*Notations:* We collect here a few standard notations used throughout the paper. Bold face small letters represent vectors. The notation $\delta = \frac{M}{N}$ denotes under-sampling ratio, and $\rho = \frac{K}{M}$ a sparsity measure. The symbol $\|x\|_2$ denotes the $L_2$ norm of the vector $x$. The operators $\mathbb{E}[X]$ and $\mathbb{V}[X]$ denote, respectively, the mean and the variance of a random variable $X$ and $\ln(x)$ denotes the natural logarithm of $x$.

## II. RESTRICTED ISOMETRY RANDOM VARIABLE

In this section, we describe the RIVs using functional mappings. In the next section, we aim to determine their PDFs.

### 1. Description

The left and right RICs in (1) are constants for a deterministic encoder. However, these RICs become random variables for random encoders. We refer to these random variables, in this paper, as *restricted isometry random variables* (RIVs). The sample space for these random variables is an ensemble of random encoders. Thus, associated with the random ensemble, there are two random variables, namely, the left RIV and the right RIV defined as follows:

***Definition 2 Restricted isometry random variables* (RIVs):** Let $\Omega = \{A_i\}_{i=1}^{\infty}$ denote an ensemble, a set of $M \times N$ random encoders. The left RIV is a function $\Delta_K^L : \Omega \to [0,1]$ which maps an encoder in the ensemble into a number between zero and one. Similarly, the right RIV is a function $\Delta_K^R : \Omega \to [0,\infty]$ that maps the encoders into number greater than zero.

We note that while the RICs characterize a single deterministic encoder, the RIVs characterize the entire ensemble. Thus, by studying the RIVs, we will be able to address the questions regarding the entire ensemble.

Gaussian ensemble is widely studied in CS. The reasons include that the Gaussian ensemble has a nice mathematical structure and as such it allows tractable analysis. Besides, Gaussian ensemble is natural and is found in real systems such as the turbid media imaging [38].

In this paper, we aim to determine the PDF of the RIVs for a Gaussian ensemble. By Gaussian ensemble, we mean a collection of $M \times N$ encoders whose entries are i.i.d. Gaussian random variables with zero mean and variance $\sigma^2$.

### 2. RIVs of Gaussian Encoders

We define the RIVs of the Gaussian encoders via the following two mappings:

*Mapping 1.* For a fixed support set, this mapping transforms the entire ensemble in to an intermediate random variable called the *ratio random variable*.

*Mapping 2.* This step transforms the ratio random variables into the RIVs via the order statistics of random variables.

*Mapping 1:*
Consider the ratio (1), which is also the Rayleigh quotient.

For a fixed support set $\mathcal{S}$ and $\boldsymbol{x} \in \Psi$ and for the random draw of $A$ from the ensemble, the ratio

$$R(\mathcal{S}) := \frac{\|A\boldsymbol{x}\|_2^2}{\|\boldsymbol{x}\|_2^2} \quad (3)$$

is a random variable. We refer to $R(\mathcal{S})$ as the *ratio random variable*, which is described in Lemma 3.

**Lemma 3:** Let $A$ be an i.i.d. Gaussian encoder whose $(i,j)$th element $a_{i,j}$ is a Gaussian random variable with zero mean and variance $\sigma^2$. For any $\boldsymbol{x} \in \Psi$, the $R(\mathcal{S})$ in (3) is a central Chi-square random variable with $M$ degrees of freedom.

Proof: We assign the numerator in (3) as $U_\mathcal{S} = \|A\boldsymbol{x}\|_2^2 = \|A_\mathcal{S} \boldsymbol{x}_\mathcal{S}\|_2^2 = \sum_{i=1}^{M} y_\mathcal{S}^2(i)$ and $c_\mathcal{S} = \frac{1}{\|\boldsymbol{x}\|_2^2} = \frac{1}{\sum_{j=1}^{K} x_{\mathcal{S}[j]}^2}$ and thus, $R(\mathcal{S}) = c_\mathcal{S} U_\mathcal{S}$. The PDF of $R(\mathcal{S})$ can be determined by finding the PDF of $U_\mathcal{S}$. In $U_\mathcal{S} = \sum_{i=1}^{M} y_\mathcal{S}^2(i)$, the $y_\mathcal{S}(i)$ denote the inner product of $i$th row of $A$ with $\boldsymbol{x}$ given by $y_\mathcal{S}(i) = \sum_{j=1}^{K} a_{i,\mathcal{S}[j]} x_{\mathcal{S}[j]}$. Now, $y_\mathcal{S}(i)$ is a Gaussian random variable with mean $\mathbb{E}[y_\mathcal{S}(i)] = 0$ and variance $\mathbb{V}[y_\mathcal{S}(i)] = \sum_{j=1}^{K} \mathbb{V}(a_{i,\mathcal{S}[j]} x_{\mathcal{S}[j]}) = \sigma^2 \sum_{j=1}^{K} x_{\mathcal{S}[j]}^2 = \frac{\sigma^2}{c_\mathcal{S}}$. Then, $y_\mathcal{S}^2(i)$ is a central Chi-square random variable with 1-degree of freedom. The collection of Chi-square random variables $\{y_\mathcal{S}^2(i)\}$ are independent, because the $M$ rows of $A$ are independent with each other. Since $U_\mathcal{S}$ is an addition of $M$ i.i.d. Chi-square random variables, it follows the Chi-square distribution with $M$-degrees of freedom. The mean of $U_\mathcal{S}$ is $\mathbb{E}[U_\mathcal{S}] = \frac{M\sigma^2}{c_\mathcal{S}}$ and its variance is $\mathbb{V}[U_\mathcal{S}] = \frac{2M\sigma^4}{c_\mathcal{S}^2}$. The random variable $R(\mathcal{S})$ is a scaled version of $U_\mathcal{S}$ with the scaling factor $c_\mathcal{S}$. Hence, $R(\mathcal{S})$ is also a Chi-square random variable with the mean $\mathbb{E}[R(\mathcal{S})] = M\sigma^2$ and $\mathbb{V}[R(\mathcal{S})] = 2M\sigma^4$. ∎

During the review period of this paper, via a private e-mail correspondence with Jared Tanner, the paper by Cartis and Thompson [39] was brought to our attention, in which Lemma 4.1 mirrors the results stated in Lemma 3. In fact, Lemma 4.1 is taken from [40], which derives the distributions of the Rayleigh quotient in (3) for the general case when the "entries of $A$ are not necessarily independent."

We provide the following remarks for Lemma 3.

**Remark 1.** It is interesting to observe that the PDF of $R(\mathcal{S})$ is identical *for each* support set $\mathcal{S}$ and independent of the values taken by $\boldsymbol{x}$. *Rationale:* Since $R(\mathcal{S})$ is a central Chi-square random variable, it is completely characterized by its degrees of freedom $M$. Note also that the mean $\mathbb{E}[R(\mathcal{S})]$ and the variance $\mathbb{V}[R(\mathcal{S})]$ depend only on $M$. Thus, the PDF of $R(\mathcal{S})$ does not depend on $\boldsymbol{x}$ nor on its support $\mathcal{S}$. It is also interesting to note that the PDF is independent of $K$.

Based on (1) and Remark 1, we state the following corollary.

*Corollary 4:* Let $\Omega$ denote the ensemble of $M \times N$ Gaussian encoders and $R(\mathcal{S})$ the ratio random variable defined in (3). Then, for every $\boldsymbol{x} \in \Psi$, the following relation holds

$$\left(1 - \delta_K^L\right) \le R(\mathcal{S}) \le \left(1 + \delta_K^R\right) \quad (4)$$

A quick comparison of (4) with (2) tells that the ratio random variable can also be used to represent the RICs. We explore this link between RICs and the ratio random variables further in Mapping 2.

**Remark 2.** In [41], Park and Lee have shown that for binary $K$-sparse signals, the mean and variance of the numerator in (3) respectively, are $K$ and $\frac{2K^2}{M}$. Their result, which is the subset of ours, can be obtained by setting $c_\mathcal{S} = \frac{1}{K}$ in Lemma 3.

**Remark 3.** Lemma 3 is for the Gaussian ensemble. For non-Gaussian ensembles, such as the Bernoulli, work needs to be done while the procedure in this paper can be applied there.

In summary, we illustrated that the ratio random variable follows the Chi-square distribution for each support set.

In mapping 2, we aim to relate the ratio random variables to the RIVs via the minimum and the maximum order statistics of the i.i.d. random variables.

*Mapping 2:*

Using Corollary 4, we define the RIVs. The left RIV, $\Delta_K^L$ is defined as the minimum of the i.i.d. ratio random variables $R(\mathcal{S}_i), i = 1, 2, \cdots, N_S$, $N_S := \binom{N}{K}$ given by

$$\Delta_K^L := 1 - \min_{\mathcal{S}_i} R(\mathcal{S}_i) \quad (5)$$

Of course, any two ratio random variables $R(\mathcal{S}_i)$ and $R(\mathcal{S}_j), i \ne j$ in (5) in general are not independent with each other due to the overlapping support sets; but, (5) with the i.i.d. assumption is sufficient for RIC prediction and phase transition analysis discussed in Section V. For more details, please see Remark 4 in Section V and Appendix E. In Appendix E, we call (5) and (6) as dependent representation of the RICs, when the random variables $R(\mathcal{S}_i)$ and $R(\mathcal{S}_j), i \ne j$ are dependent, otherwise we call them as i.i.d. representations of the RICs. In the subsequent discussions, we assume the i.i.d. representation. The left RIV characterizes the probabilistic nature of the left RIC. Similarly, we define the right RIV, $\Delta_K^R$ as the maximum among the i.i.d. ratio random variables as

$$\Delta_K^R := \max_{\mathcal{S}_i} R(\mathcal{S}_i) - 1 \quad (6)$$

The right RIV characterizes the probabilistic nature of the right RIC. In this paper, by *RIV approach*, we mean the representation of the RICs via the ratio random variables in (5) and in (6) and their subsequent prediction via the critical functions in Section V. Since the PDF of $R(\mathcal{S})$ is known, it is possible to find the PDF of the RIVs in (5) and (6). With these definitions for the RIVs, we are in a position to determine their PDFs.

*3. Non-asymptotic and asymptotic distributions*

In the subsequent sections, we aim to determine the distributions (PDF and CDF) of the RIVs for two different scenarios, namely, *non-asymptotic and asymptotic*. In this subsection, we describe the convergence phenomenon that distinguishes these two scenarios.

The RIVs in (5) and (6) are defined via the minimum and the maximum order statistics of $N_S$ number of i.i.d. Chi-square random variables. Hence, the CDF of the RIVs can be derived straightforwardly from these definitions (Appendices B and C).

However, when $N_S$ is sufficiently large, the CDF of the order statistics of the i.i.d. random variables converges (in distribution) to the extreme value distributions such as the Weibull or the Gumbel [42]. This convergence is extensively studied in the extreme value theory for several types of random variables including the Chi-square (More details are given in Appendix D). Since RIVs are defined via the order statistics, the CDF of the RIVs also converges to the extreme value distributions. We find it useful to define two scenarios where the convergence occurs and where it does not. It is known [43, Theorem 3.2] that the convergence rate of the order statistics of the Chi-square (special case of Gamma) random variables is $\mathcal{O}(\log N_s)^{-2}$. Hence, we devise a rule for testing the convergence based on the value of $(\log N_s)^{-2}$.

We refer to the scenarios where $(\log N_s)^{-2} < \eta$ as asymptotic and the corresponding value of $N_S$ as a sufficiently large value, $\eta > 0$ is a small number. When $(\log N_s)^{-2} > \eta$, we refer to the scenario as *non-asymptotic*. In typical CS systems such as the turbid lens imaging [38], the problem sizes are $N = 2 \times 10^4$ and $K = 147$ and hence, $(\log N_s)^{-2}$ is a very small value. These systems fall within the asymptotic category that invites us to explore the RIVs for this category.

An approach similar to the RIV (only in the sense of using the ratio random variable for describing the RICs) is the *proportional growth asymptotic* framework by Cartis and Thompson [42] for studying the RIC-based phase transition analysis of greedy recovery algorithms. The RIV framework and proportional growth asymptotic framework provide similar results for asymptotic problem sizes. The RIV framework, however, is general in that it includes both the asymptotic and the finite-size problems. Moreover, RIV framework is much easier to handle due to the availability of the precise PDF and CDF of the RIVs as well as the closed-form critical functions.

### III. NON-ASYMPTOTIC DISTRIBUTIONS OF RIVS

In this section, we derive the CDF and the PDF of the RIVs for the non-asymptotic case. Asymptotic scenarios are discussed in Section IV.

*1. Distributions of the left RIV*

The non-asymptotic distributions of the left RIV are stated in the following theorem.

***Theorem 5 CDF and PDF of the left RIV:*** Let $\Omega = \{A_1, A_2, \cdots\}$ denote an ensemble of $M \times N$ Gaussian encoders and $\Delta_K^L$ be the left RIV defined in (5). The non-asymptotic CDF of the left RIV for $0 \leq u \leq 1$ is given by

$$F_L(u) = \left[1 - \frac{\gamma\left(\frac{M}{2}, \frac{M(1-u)}{2}\right)}{\Gamma\left(\frac{M}{2}\right)}\right]^{N_S} \qquad (7)$$

and the corresponding non-asymptotic PDF is given by

$$p_L(u) = \frac{N_S \left(\frac{M}{2}\right)^{\frac{M}{2}} e^{-\frac{M}{2}}}{\Gamma\left(\frac{M}{2}\right)} \left[1 - \frac{\gamma\left(\frac{M}{2}, \frac{M(1-u)}{2}\right)}{\Gamma\left(\frac{M}{2}\right)}\right]^{N_S - 1} (1-u)^{\frac{M}{2}-1} e^{\frac{Mu}{2}}. \qquad (8)$$

where $\gamma(.,.)$ is the lower incomplete gamma integral [44, Eq. 3.381.1] and $\Gamma(.)$ is the gamma function.

Proof: Please see Appendix B. ∎

The CDF and the PDF in Theorem 5 are, as expected, functions of the triplets $(K, M, N)$. While the dependence on $M$ arrives solely from the ratio random variable, their dependence on $N$ and $K$ comes from the total number of support sets $N_S$.

*2. Distribution of the right RIV*

We state the CDF and the PDF of the right RIV in the following theorem.

***Theorem 6 CDF and PDF of the right RIV:*** Let $\Omega = \{A_1, A_2, \cdots\}$ denote an ensemble of $M \times N$ Gaussian encoders and $\Delta_K^R$ be the right RIV defined in (6). The non-asymptotic CDF of the right RIV is given by

$$F_R(v) = \left[\frac{\gamma\left(\frac{M}{2}, \frac{M(v+1)}{2}\right)}{\Gamma\left(\frac{M}{2}\right)}\right]^{N_S} \quad v \geq 0 \qquad (9)$$

and the corresponding non-asymptotic PDF is

$$p_R(v) = \frac{N_S \left(\frac{M}{2}\right)^{\frac{M}{2}} e^{-\frac{M}{2}}}{\Gamma\left(\frac{M}{2}\right)} \left[\frac{\gamma\left(\frac{M}{2}, \frac{M(v+1)}{2}\right)}{\Gamma\left(\frac{M}{2}\right)}\right]^{N_S - 1} (v+1)^{\frac{M}{2}-1} e^{-\frac{Mv}{2}}. \qquad (10)$$

Proof: Please see Appendix C. ∎

In Fig. 1, we plot the PDF of the left RIV in (8) and that of the right RIV in (10) for $(5, 200, 1000)$. We note that the PDFs of the RIVs are asymmetric with a sharp tail on the lower end. We also observe that the range of sample values of the left RIV is always confined between zero and one and while that of the right RIV can extend beyond one.

## IV. ASYMPTOTIC DISTRIBUTIONS OF RIVS

In this section, we derive the asymptotic PDF of the RIVs. As mentioned in Section II.3, when $N_S$ is sufficiently large, the CDF of the order statistics converges to the extreme value distributions. It is well-known [42] that the minimum order statistics of large number of i.i.d. Chi-square random variables converges to the Weibull distribution (see Appendix D. 1). In this paper, since we defined the left RIV in terms of the minimum order statistics, we show (in Lemma 14) that the left RIV follows the Weibull distribution. In addition, the parameters of the Weibull distribution such as the location, scale and shape constants pertaining to the left RIV are derived in Appendix D, which is one of our new contributions in this paper. Similarly, we show, in Lemma 15, that the right RIV follows the Gumbel distribution.

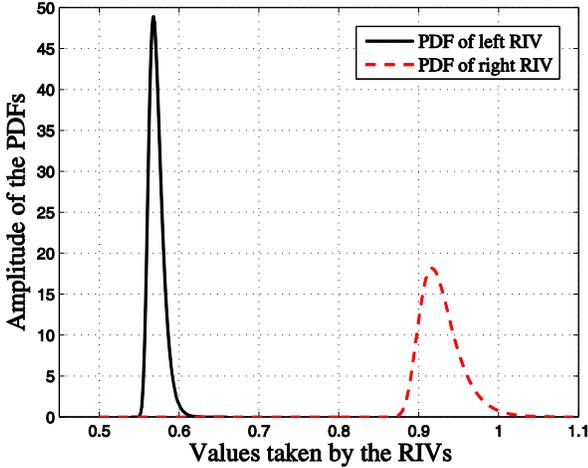

Fig. 1. The PDF of left and right RIVs for the triplet $(5, 200, 1000)$.

*1. Distribution of the left RIV: Weibull*

In what follows, we state the asymptotic PDF of the left RIV.

**Theorem 7 Asymptotic CDF and PDF of the left RIV:** Let $\Omega$ denote an ensemble of $M \times N$ Gaussian encoders and $\Delta_K^L$ be the left RIV in (5). The asymptotic CDF of the left RIV is a Weibull distribution and it is given by

$$F_L^\infty(u) = \exp\left(-\left(\frac{1-u}{q}\right)^\beta\right) \quad 0 \leq u \leq 1 \quad (11)$$

where the constants $q = \frac{6}{M}\exp\left(\frac{2}{M}\left[\ln\frac{M}{2} + \ln\Gamma\left(\frac{M}{2}\right) - \ln N_S\right]\right)$ and $\beta = \frac{M}{2}$. The corresponding asymptotic PDF is given by

$$P_L^\infty(u) = \frac{\beta}{q}\left(\frac{1-u}{q}\right)^{\beta-1}\exp\left(-\left(\frac{1-u}{q}\right)^\beta\right) \quad (12)$$

Proof: Please see Appendix D.1 ∎

Rather than plotting the asymptotic PDF, it would be interesting to plot its *support*. This kind of plotting is useful, because, from a single plot we can quickly ascertain the range of values that the RIV takes for various triplets.

The non-zero *support of a PDF* is defined as an interval of points within which the PDF remains non-zero, or sufficiently away from being zero. We refer to the starting point of the interval as a *lower-end non-zero support point* (LESP) and the ending point of the interval as an *upper-end non-zero support point* (UESP). The numerical difference between the UESP and the LESP is then the *width of the support*. The support of the asymptotic PDF of the left RIV is given in the following Lemma.

**Lemma 8 (Support of the pdf of the left RIV):** Let $P_L^\infty(u)$ be the asymptotic PDF of the left RIV given in (12) with the parameters $q$ and $\beta$. Then, the support of the PDF (or its CDF) is given by the set

$$S_L = \left\{u \in [0,1]: \left(1-q\left[\ln\tfrac{1}{\varepsilon}\right]^{\tfrac{1}{\beta}}\right) \leq u \leq \left(1-q\left[\ln\tfrac{1}{1-\varepsilon}\right]^{\tfrac{1}{\beta}}\right)\right\}$$

where $\varepsilon > 0$ is a very small number.

Proof: Please see Appendix D.2. ∎

In Fig. 2, we plot the support of the PDF of the left RIV for various $\rho$ and $\delta$. We plot the graph by considering $\varepsilon = 1 \times e^{-6}$, $M = 1000$, $N = 5 \times 10^4$ and we vary $N$ from $N$ to $M$. In Fig. 2, the vertical distance between the two adjacent curves (of the same line style) indicates the width of the support. For instance, for $\delta = 0.2$ and $\rho = 0.05$, the value 0.397 on the top curve indicates the UESP of the support, and the value 0.354 on the bottom curve dictates the LESP of the support. The width of the support (the patch) is 0.043. We note from Fig. 2 that as $\delta$ increases, the support moves towards zero. This behavior of the left RIV agrees with the fact that the RICs of those encoders that undersample a signal only slightly are close to zero.

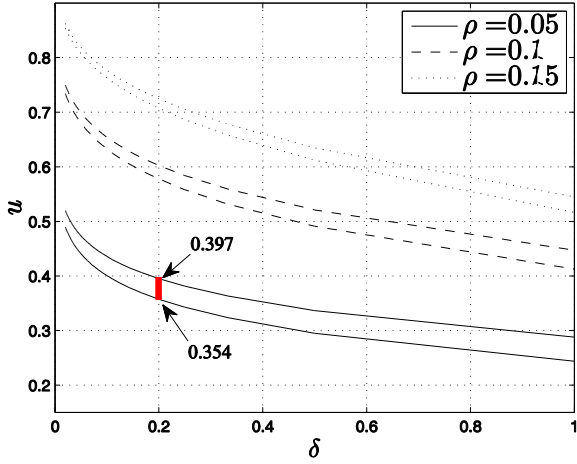

Fig.2. Support of the PDF of the left RIV.

*2. Distribution of the right RIV: Gumbel*

Similar to the left RIV, we present, in this section, the asymptotic distribution of the right RIV.

***Theorem 9 Asymptotic CDF and PDF of right RIV (Gumbel):*** Let $\Omega = \{A_1, A_2, \cdots\}$ denote an ensemble of $M \times N$ Gaussian encoders and $\Delta_K^R$ be the right RIV defined in (6). The asymptotic CDF of the right RIV for $v \geq 0$ is a Gumbel distribution and it is given by

$$F_R^\infty(v) = \exp\left(-\exp\left(-\left(\tfrac{v+1-l}{s}\right)\right)\right) \quad (13)$$

where $s = \tfrac{2}{M}$ is the scaling constant of the distribution and $l = s \cdot \left(\ln N_S + \left(\tfrac{M}{2}-1\right)\ln \ln N_S - \ln \Gamma\left(\tfrac{M}{2}\right)\right)$ is the location constant. The corresponding asymptotic PDF is given by

$$P_R^\infty(v) = \frac{1}{s}\exp\left(-\left[\exp\left(-\tfrac{v+1-l}{s}\right) + \left(\tfrac{v+1-l}{s}\right)\right]\right) \quad (14)$$

Proof: Please see Appendix D.3. ∎

As done for the left RIV, the support of the PDF of the right RIV can be easily defined and plotted using the distributions in Theorem 9. Due to space limitations, however, we skip those details.

V. ROLE OF RIVS IN RIC PREDICTION AND PHASE TRANSITION ANALYSIS

In the previous sections, we derived the PDF of the RIVs. In this section, we elaborate on the benefit of these PDFs for predicting the RICs and for analyzing the PT of $L_1$ decoders.

*1. RIC Prediction*

RIC prediction deals with determining the values of the RICs of an encoder. For a Gaussian encoder, RIC prediction is typically done [6] by using critical functions. In particular, the level curves of these functions in the PT space provide an upper bound on the RICs that an encoder is very unlikely to exceed. That is, the area below a level curve designates a region in which encoders with RICs less than the level value are guaranteed to exist with overwhelming probability. Thus, critical functions play a major role in RIC prediction.

Candes and Tao [1] and Blanchard *et al.* [6] derived the critical functions using the EV approach. In Subsection V.3, we aim to derive new critical functions by adopting the RIV approach. We illustrate that the new critical functions significantly improve the RIC prediction much better than that of the EV approach. Before we discuss about the new critical functions, it will be helpful to spend some time in highlighting the differences between the RIV and the EV approaches. This will help us to understand better where the prediction improvement for the RIV approach originates from.

*2. RIV and EV approaches: A comparison*

Blanchard *et al.* represent, respectively, the left and the right RICs [6, p. 6] as

$$L_K = 1 - \min_{\mathcal{S}_i} \lambda_{\min}\left(A_{\mathcal{S}_i}^T A_{\mathcal{S}_i}\right) \quad (15)$$

and

$$U_K = \max_{\mathcal{S}_i} \lambda_{\max}\left(A_{\mathcal{S}_i}^T A_{\mathcal{S}_i}\right) - 1 \quad (16)$$

where $\lambda_{\min}\left(A_{\mathcal{S}_i}^T A_{\mathcal{S}_i}\right)$ and $\lambda_{\max}\left(A_{\mathcal{S}_i}^T A_{\mathcal{S}_i}\right)$ denote the smallest and the largest eigenvalues of the Wishart matrix $A_{\mathcal{S}_i}^T A_{\mathcal{S}_i}$, respectively. Equations (15) and (16) provide a way to predict the RICs by using the PDF of the eigenvalues. Interesting facts emerge when we compare (5) and (6) with (15) and (16). These facts are given below. For brevity, we drop the arguments of $\lambda_{\min}, \lambda_{\max}$, and $R$, in the subsequent discussions.

1. <u>Differences in formulation:</u> The RIVs in (5) and (6) define the RICs via the order statistics of $R$, a Chi-square random variable. On the other hand, in the EV approach ((15) and (16)), the RICs are defined in terms of the order statistics of the eigenvalues. Thus, the RIV and the EV approaches are different in their formulation.
2. <u>Differences in modeling the RIC:</u> In the RIV approach, a probability measure is induced on the RICs via the PDF of $R$. Whereas in the EV approach, the RICs are induced with the PDF of the eigenvalues. Thus, the RIV and the EV approaches compute the RIC values using two different sets of PDFs. Now the question is which one among the two sets models the RICs more accurately? We address this question via an example.

   In Fig. 3, we plot the PDFs of the ratio $R$, $\lambda_{\min}$ and $\lambda_{\max}$ [46] for a fixed support set. To plot the graphs, we set $M = 700$ and $K = 55$. From Fig. 3, we note that $1-\delta_K^L$ denotes the LESP of the PDF of the ratio and $1-L_K$, the LESP of the PDF of $\lambda_{\min}$. Judging from Fig. 3, $1-L_K$ is smaller than $1-\delta_K^L$. In other words, $L_K$ is greater than $\delta_K^L$.

Thus, $L_K$ over predicts the RIC $\delta_K^L$, which is the true left RIC. From Fig. 5 we read $1-\delta_K^L = 0.8$ and hence $\delta_K^L = 0.2$. On the other hand, $1-L_K = 0.47$ or the RIC value predicted by the EV approach is $L_K = 0.53$.

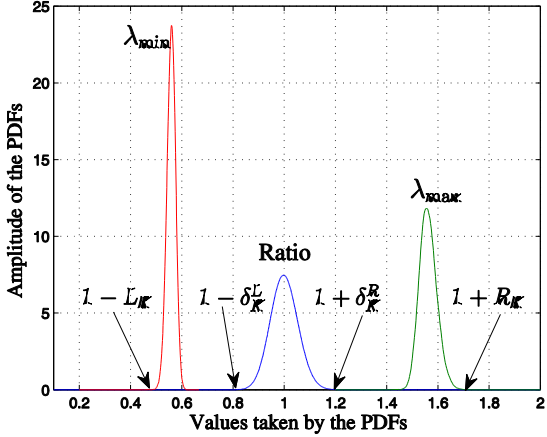

Fig. 3. PDFs of eigenvalues and the ratio for $M = 700$ and $K = 55$.

This example shows that the left RIC predicted by the eigenvalues is greater than that of by the ratio. Similarly, it can be shown that the eigenvalues over predict the right RIC as well. Thus, the RIC values predicted by the EV approach act as loose upper bounds for the true RIC values predicted by the RIV approach. The above example is not an isolated case, and in fact, it is true that the ratio being the Rayleigh quotient obeys the following well-known inequality:

$$\lambda_{\min}\left(A_{\mathcal{S}}^T A_{\mathcal{S}}\right) \leq R(\mathcal{S}) \leq \lambda_{\max}\left(A_{\mathcal{S}}^T A_{\mathcal{S}}\right). \quad (17)$$

While the EV approach accesses the RICs via the eigenvalues, the RIV approach accesses them via the ratio random variables. Therefore, the RIV approach provides an accurate picture of the RICs.

Thus, from the discussion above, it is clear that the RIC prediction via the RIV is superior to that of via the EV approach. We are now in a position to derive the critical functions.

*3. Critical functions for the left and the right RICs*

The derivation of the critical functions begins with the representations for the RICs, which are given in (5) and (6) for the RIV and in (15) and (16) for the EV approach. In EV approach, the critical function for the left RIC, denoted as $\mathcal{L}(\rho,\delta)$, is derived in [6] such that $\Pr\{L_K \leq \mathcal{L}(\rho,\delta)\} \approx 1-\varepsilon$, where $\varepsilon$ is a very small number. In a similar manner, the critical function for the right RIC, $\mathcal{U}(\rho,\delta)$, is also derived [6, Theorem 2.3].

In this subsection, we aim to derive new critical functions for the RICs via the RIV approach. We denote the critical function for the left RIC as $u(\rho,\delta)$ and that of for the right RIC as $v(\rho,\delta)$. We find $u(\rho,\delta)$ such that $\Pr\{\Delta_K^L \leq u(\rho,\delta)\} = F_L^\infty(u(\rho,\delta)) = 1-\varepsilon$. In a similar manner, we derive the critical function for the right RIC. Since we have a closed form expression for the CDF $F_L^\infty(u)$, the critical functions can be readily derived.

*Remark 3:* For deriving the new critical functions, we make use of the asymptotic CDF in (11), in order to be consistent with the EV-based proportional-growth asymptotic approach in [1] and [6]. In addition, the realistic problem sizes often fall into the asymptotic case, say, $N$ on the order of millions and $K$ on the order of a few thousands.

*Lemma 10 (Critical function for the left RIC):* Let $\Delta_K^L$ be the left RIV defined in (5) with the CDF $F_L^\infty(u) = \Pr\{\Delta_K^L \leq u(\rho,\delta)\}$. Let $\varepsilon$ be a very small number, $K = \rho M$, $M = \delta N$ and $(\rho,\delta) \in [0,1]^2$. The critical function $u(\rho,\delta)$ for the left RIC is then given by

$$u(\rho,\delta) = 1 - \tfrac{6}{\delta N}\exp\left(\tfrac{2}{\delta N}\left[\ln\tfrac{\delta N}{2} + \ln\Gamma\left(\tfrac{\delta N}{2}\right) - \ln\binom{N}{\delta\rho N}\right]\right)\left[\ln\left(\tfrac{1}{1-\varepsilon}\right)\right]^{\tfrac{2}{\delta N}} \quad (18)$$

Proof: Setting $F_L^\infty(u) = 1-\varepsilon$ and solving for $u$ returns a function in terms of $(K, M, N)$. By substituting $K = \rho\delta N$ and $M = \delta N$ in the resulting function, we obtain (18). ∎

We compare the RIC values predicted by $u(\rho,\delta)$ with that of $\mathcal{L}_C(\rho,\delta)$ by Candes and Tao [22, p. 2889] and $\mathcal{L}(\rho,\delta)$ by Bah *et al.* [22]. For this purpose, we plot, in Fig. 4, the level curves of these functions for the level value of 0.6, $\varepsilon = 1 \times e^{-3}$, and $N = 10^4$. The level curves of $\mathcal{L}_C(\rho,\delta)$ and $\mathcal{L}(\rho,\delta)$ are obtained from [47]. We recall from Section V.1 that in the region below the level curves, Gaussian encoders are guaranteed to exist with probability 1. We refer to this region as RIC *predicted region*. A critical function that predicts a larger portion of the RIC predicted region is deemed good. The reason is that large RIC predicted region implies the existence of large number of Gaussian encoders with probability 1.

From Fig. 4, we observe that the level curve of the RIV critical function predicts a larger RIC predicted region than the EV-based critical functions. We cross check the region predicted by the RIV with the results shown in Fig. 2, which shows the behavior of the left RIC for various $\rho$ and $\delta$. Let us suppose that $\rho = 0.1$. In Fig. 2, at this value of $\rho$, the values taken by the left RIC are less than 0.6 for any $\delta \geq 0.2$. For $\delta < 0.2$, the left RIC values exceed 0.6. Let us examine Fig. 4 at $\rho = 0.1$. We observe that at this value of $\rho$, the RIC values are predicted to exceed the level value of 0.6 only when $\delta < 0.2$, which we have already confirmed using Fig. 2. The reason that the RIV approach achieves a larger RIC prediction region than the EV approach is due to the precise representation of the RICs using the order statistics of the ratio random variable rather than using the eigenvalues.

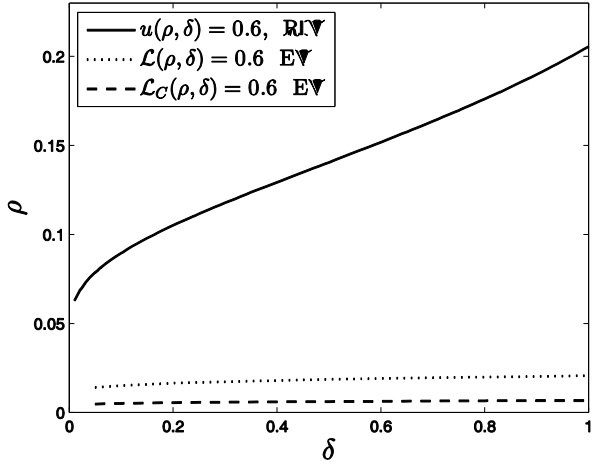

Fig. 4. Prediction of the left RIC by various approaches

Similarly, for the right RIC, we state the critical functions in the following Lemma.

**Lemma 11 (Critical function for the right RIC):** Let $\Delta_K^R$ be the right RIV defined in (6) with the CDF $F_R^\infty(v) = \Pr\{\Delta_K^R \leq v(\rho,\delta)\}$. Let $\varepsilon$ be a very small number, $K = \rho M$, $M = \delta N$, and $(\rho,\delta) \in [0,1]^2$. The critical function for the right RIC is then given by

$$v(\rho,\delta) = \tfrac{2}{\delta N}\left[\ln\binom{N}{\rho\delta N} + \left(\tfrac{N\delta}{2}-1\right)\ln\ln\binom{N}{\rho\delta N} - \ln\Gamma\delta N\right] \\ -1 - \tfrac{2}{\delta N}\ln\ln\left(\tfrac{1}{1-\varepsilon}\right). \quad (19)$$

Proof: Setting $F_R^\infty(v) = 1-\varepsilon$ and solving for $v$ returns a function in terms of $(K,M,N)$. By substituting $K = \rho\delta N$ and $M = \delta N$ in the resulting function, we obtain (19). ∎

We report here that we obtained a similar RIC prediction improvement for the right RIC as that of for the left RIC in Fig. 4. Due to space limitations, we skip those discussions.

*Remark 4:* We used the i.i.d. representation of the RICs ((5) and (6)) for RIC prediction and we have a proof in the Appendix E that this representation is sufficient for the RIC prediction. The key idea in the proof is to show that the level curve (in Fig 4) and phase boundary (in Fig. 5) obtained using the i.i.d. representation are below those of obtained by using the dependent representation. For this purpose, we show that it is enough to prove the CDF of the RIV (left and right) using the dependent representation is always greater than or equal to the CDF of the RIV i.i.d. representation. Please see Appendix E for more details.

*4. RIVs in phase transition analysis*

PT analysis is a standard framework in CS for studying the tradeoff between the signal sparsity and the undersampling [28]. Given an encoder/decoder pair, PT analysis aims to identify a boundary that separates the region of successful signal recovery from the failure recovery region.

In this section, we aim to study the PT boundaries for the Gaussian encoder and $L_1$ decoder pair. Towards this end, we define a region in the PT space where the $L_1$ decoders guarantee perfect recovery of sparse signals. We refer this region as the *region of strong equivalence* [6]. Our aim is to find the PT boundary that identifies the region of strong equivalence.

Our approach can be easily extended to other decoders such as the $l_q$ minimization [3], the orthogonal matching pursuit (OMP) [48], and the regularized OMP [49].

***Definition 12: Strong equivalence region [6]:*** For a Gaussian ensemble $\Omega$ and an $L_1$ decoder, the *region of strong equivalence* is the region in the PT space such that the decoder exactly recovers *every K*-sparse signal $x$ from $y = Ax$.

In what follows, we briefly discuss a few approaches for identifying the region of strong equivalence. We aim not to dwell too much into technical details rather we briefly discuss their principles.

*Polytope approach (Donoho and Tanner [24]):*

This is a geometric approach. Polytopes are geometric objects having faces in many dimensions. The number of faces of a polytope can be counted. In general, a K-dimensional face of a polytope is counted using the combinatorial geometry. Face counts of the polytopes have intriguing connections with the problem of recovering sparse signals using the $L_1$ decoders.

In polytope approach, a K-sparse signal is considered as a signal-polytope (SP). The measurement is considered as a projected polytope (PP); the projector here is an encoder. Let $f_K(\text{SP})$ and $f_K(\text{PP})$ denote the K-dimensional face counts of the SP and the PP. Then, Donoho and Tanner showed that

$$\Pr\{\text{Exact Recovery using } L_1\} = \frac{f_K(\text{PP})}{f_K(\text{SP})} \quad (20)$$

That is, the ratio between the face counts of the projected polytope to that of the signal polytope is equal to the recovery probability of the $L_1$ decoder. Ideally, one aims to derive the conditions under which $f_K(\text{PP}) = f_K(\text{SP})$. This, in technical terms, is called *an agreement between the two face counts*. Using an upper bound for the ratio in (20), Donoho and Tanner showed that

$$\lim_{N\to\infty} \frac{f_K(\text{PP})}{f_K(\text{SP})} = \begin{cases} 1, & \rho < \rho(\delta,Q) \\ 0, & \rho > \rho(\delta,Q) \end{cases}.$$

That is, the $L_1$ decoder recovers the signals with probability 1, if $\rho < \rho(\delta,Q)$, where $\rho(\delta,Q)$ is the PT boundary for a particular signal polytope $Q$. It is worth to mention that Donoho and Tanner identify two kinds of phase transitions, namely, the *strong* and the *weak*, based on the strong or the weak agreement between the face counts. They are defined below.

Weak phase transitions:

$$\lim_{N\to\infty} \frac{E[f_K(\text{PP})]}{f_K(\text{SP})} = \begin{cases} 1, & \rho < \rho_w(\delta,Q) \\ 0, & \rho > \rho_w(\delta,Q) \end{cases},$$

where $E[f_K(\text{PP})]$ denotes the expected number of face counts.

Strong phase transitions:

$$\lim_{N \to \infty} \Pr(f_K(\text{PP}) = f_K(\text{SP})) = \begin{cases} 1, & \rho < \rho_S(\delta, Q) \\ 0, & \rho > \rho_S(\delta, Q) \end{cases}$$

Thus, the region of strong equivalence using the polytope approach is the region below the strong PT boundary $\rho_S(\delta, Q)$. In this paper, we compare our results with this strong PT boundary (see Fig. 5). The programs for duplicating the strong and the weak PT boundaries are available in [50].

**Geometric function analysis (Rudelson and Vershynin [20]):**

Geometric functional analysis (GFA) is yet another geometric approach different from the polytope approach. In this approach, the exact reconstruction of the sparse signals is (geometrically) viewed as embedding of cones that are missed by the kernel (null-space) of an encoder $A$. The tightness of the embedding is attained by employing the Gordon's "escape through the mesh" theorem on the kernel of $A$. This results in the PT boundary $\rho_S^{RV}(\delta)$ which is the solution to the equation $\mu^{RV}(\rho, \delta) = 1$ [22, Theorem 4.1], [6, Definition 3.8], where

$$\mu^{RV}(\rho, \delta) := \rho \left(12 + 8 \log\left(\tfrac{1}{\rho\delta}\right) \gamma^2(\rho\delta)\right) \text{ and}$$

$$\gamma(\rho\delta) := \exp\left(\frac{\log\left(1 + 2\log\left(\tfrac{e}{\rho\delta}\right)\right)}{4\log\left(\tfrac{e}{\rho\delta}\right)}\right). \quad (21)$$

*EV approach (Blanchard et al. [6])*

This approach is based on the RIP. In this approach, the region of strong equivalence is identified by using RIC-based sufficient conditions for signal recovery [6][27]. A typical RIC-based condition (for a particular decoder) has the following form: "if the RICs of an encoder $A$ are appropriately bounded, then the decoder exactly recovers every $K$-sparse signal." This statement apparently links the RICs of an encoder with the region of strong equivalence. For the Gaussian encoders, the RICs are predicted using the critical functions and hence, these functions can be used to derive the analytical PT boundary. By substituting the critical functions of the encoder in the RIC conditions, a bound for the region of strong equivalence for various decoders can be readily obtained. In this section, we aim to find the analytical PT boundary using the critical functions of the RIV and the EV approaches and compare them with the two geometric approaches. For this purpose, we consider the modern RIC-based recovery condition for the $L_1$ decoder given by Foucart and Lai [3].

*Lemma 13 (Foucart and Lai [3]):* For an $M \times N$ encoder with the RIC constants $\delta_{2K}^L$ and $\delta_{2K}^R$, consider the Foucart and Lai (FL) function

$$\mu^{FL} := \frac{1+\sqrt{2}}{4}\left(\frac{1+\delta_{2K}^R}{1-\delta_{2K}^L} - 1\right) \quad (22)$$

If $\mu^{FL} < 1$, then an $L_1$ decoder recovers every $K$-sparse signal.

The set of points $(\rho, \delta)$ in the PT space that satisfies the condition $\mu^{FL} < 1$, defines the region of strong equivalence for the RIP-based method. This condition can be transformed in terms of the critical function of the RIVs as

$$\mu^{RIV}(\rho, \delta, N) := \frac{1+\sqrt{2}}{4}\left(\frac{1+v(2\rho, \delta)}{1-u(2\rho, \delta)} - 1\right) \quad (23)$$

Then, the condition $\mu^{RIV}(\rho, \delta) = 1$ provides the PT boundary $\rho_S^{RIV}(\delta)$ for the RIV approach. Similarly, we can obtain the PT boundary $\rho_S^{EV}(\delta)$ for the EV approach by substituting the EV-based critical functions in (22).

In Fig. 5, we compare the PT boundaries of various approaches. The region below each boundary is the region of strong equivalence. Undoubtedly, the polytope approach predicts the largest region of strong equivalence. This is not surprising as the PT boundary from the polytope approach is derived by making precise "if and only if" formulation of the strong equivalence in terms of randomly projected polytope; see [11, Section 4] for more details.

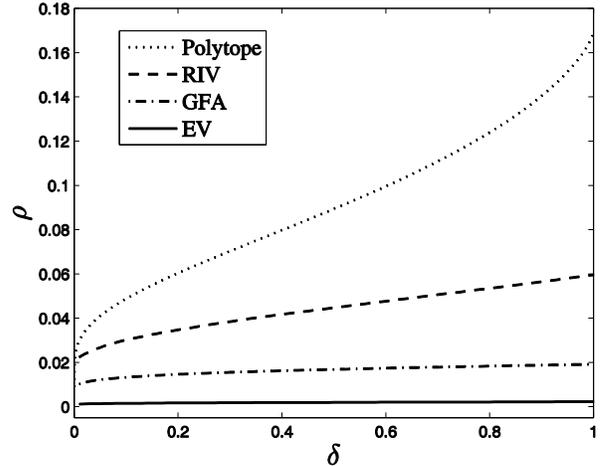

Fig. 5. Phase transitions of $L_1$ decoder by various approaches.

Interestingly, the RIV approach comes next in the line and it is more superior to the EV approach and the GFA. One way to illustrate the superiority of the RIV approach is by obtaining the sufficient number of measurements that guarantees the region of strong equivalence. We do this by using the PT boundaries in Fig. 5. The procedure is as follows [6]: in the region of strong equivalence, the values of $K$ and $M$ are such that $\frac{K}{M} \leq \rho(\delta)$ and hence, $M \geq \frac{K}{\rho(\delta)} := cK$. The constant of proportionality $c = \frac{1}{\rho(\delta)}$ can be obtained by finding the minimum value of the inverted PT boundary. The constant is listed in Table 1 for various approaches.

TABLE 1.
MEASUREMENT BOUND FROM THE PT BOUNDARIES

| Approach | Type | Measurement bound |
|---|---|---|
| Polytope | Geometric | $M \geq 5.9K$ |
| RIV | RIP-based | $M \geq 16.8K$ |
| GFA | Geometric | $M \geq 56K$ |
| EV | RIP-based | $M \geq 317K$ |

From Table 1, we note that for a given $K$, the predicted number of measurements needed is lesser for the RIV approach than the EV approach and the GFA. In particular, the RIV is 18.8 times better than the EV approach and 3 times better than the GFA. Table 1 also reveals that among the RIP-based approaches, the RIV approach is the best. It was previously believed that RIP-based approach is the least to the geometric approaches in predicting the PT boundary [6, Fig. 3.1]. However, our analysis reveals that RIP approach based on the RIV is better than that of the GFA.

## VI. CONCLUSIONS

In this paper, we introduced the concept of restricted isometry random variables (RIVs) as a generalization of the restricted isometric constants (RICs). The RIVs arise naturally in the analysis of compressive sensing (CS) systems when the encoder is random. For the widely employed Gaussian encoders, we derived the precise distribution functions of the RIVs for non-asymptotic and asymptotic scenarios.

On the application side, we elaborately discussed the benefits of the distributions for predicting the RICs and for analyzing the phase transitions (PT). Currently available critical functions for RIC prediction are based on the eigenvalue (EV) approach. We demonstrated, however, that the critical functions we derived by using the RIV approach help to improve the RIC prediction substantially. We illustrated that this improvement is possible, due to the precise formulation of the RICs via the Rayleigh quotient rather than the eigenvalues.

We investigated the role of RIVs in PT analysis. In particular, we addressed the problem of finding a region in the PT space where an $L_1$ decoder guarantees exact sparse signal recovery. We compared our results with the two geometric methods namely, the polytope and the geometric function analysis (GFA) as well as with the EV approach. We found that the RIV approach is much better than the EV approach and the GFA in terms of predicting the PT boundary and the number of measurements required for signal recovery.

## APPENDIX A
## PDF AND CDF OF CHI SQUARE RADOM VARIABLES

The PDF of a central Chi-square random variable with $M$ degrees of freedom with $\sigma^2 = \frac{1}{M}$ is

$$p_C(x) = \begin{cases} \frac{\left(\frac{M}{2}\right)^{\frac{M}{2}}}{\Gamma\left(\frac{M}{2}\right)} x^{\frac{M}{2}-1} e^{-\frac{Mx}{2}} & x > 0 \\ 0 & \text{otherwise} \end{cases} \quad (24)$$

and its corresponding CDF is given by

$$F_C(x) = \frac{\gamma\left(\frac{M}{2}, \frac{Mx}{2}\right)}{\Gamma\left(\frac{M}{2}\right)} \quad x \geq 0. \quad (25)$$

where $\gamma(.,.)$ is the lower incomplete gamma integral [41, Eq. 3.381.1].

## APPENDIX B
## PROOF OF THEOREM 5

We set $X = \min_{\mathcal{S}_i} R(\mathcal{S}_i)$, then, $\Delta_K^L = 1 - X$. The CDF of $\Delta_K^L$ for $0 \leq u \leq 1$ is written as

$$\begin{aligned} F_L(u) &= \Pr\left(\Delta_K^L \leq u\right) \\ &= 1 - F_X(1-u) \end{aligned} \quad (26)$$

The PDF of $\Delta_K^L$ is then given by $p_L(u) = \frac{d}{du} F_L(u) = p_X(1-u)$. We now obtain $p_X(u)$, by first finding its CDF $F_X(u)$. Using the CDF of the minimum of $N_S$ i.i.d. random variables [51, Eq. 2.2.11], we obtain

$$F_X(u) = 1 - \left[1 - F_C(u)\right]^{N_S}, \quad (27)$$

where $F_C(u)$ is given in (25). Then, $p_X(u)$ is $p_X(u) = \frac{d}{du} F_X(u) = N_S \left[1 - F_C(u)\right]^{N_S - 1} p_C(u)$. Now, $p_L(u) = p_X(1-u) = N_S \left[1 - F_C(1-u)\right]^{N_S - 1} p_C(1-u)$, where $p_C(u)$ is given in (24). After substituting $F_C(u)$ in $F_X(u)$ and then substituting $F_X(u)$ in (26), we arrive at the CDF in Theorem 5. Similarly, by substituting, $p_C(u)$ and $F_C(u)$ in $p_L(u)$ we obtain the PDF.

## APPENDIX C
## PROOF OF THEOREM 6

We set $Y = \max_{\mathcal{S}_i} R(\mathcal{S}_i)$, then $\Delta_K^R = Y - 1$. The CDF of $\Delta_K^R$ for $v \geq 0$ is given by

$$\begin{aligned} F_R(v) &= \Pr\left(\Delta_K^R \leq v\right) \\ &= F_Y(v+1) \end{aligned} \quad (28)$$

The PDF $p_R(v)$ of $\Delta_K^R$ is then calculated as $p_R(v) = \frac{d}{dv} F_R(v) = p_Y(v+1)$. We now find the PDF of $Y$ by first finding its CDF $F_Y(v)$. Since $Y = \max_{\mathcal{S}_i} R(\mathcal{S}_i)$, the CDF of the maximum of $N_S$ i.i.d. random variables for $y \geq 1$ $(y = v+1)$ is given by [51, Eq. 2.2.12]

$$F_Y(y) = F_C(y)^{N_S}. \quad (29)$$

The PDF $p_Y(y)$ is then determined as $p_Y(y) = \frac{d}{dy} F_Y(y) = N_S [F_C(y)]^{N_S-1} p_C(y)$. The PDF of $\Delta_K^R$, $p_R(v)$, is then obtained in terms of $p_C(y)$ as and $F_C(y)$ as $p_R(v) = p_Y(v+1) = N_S [F_C(v+1)]^{N_S-1} p_C(v+1)$. The result of substituting $F_C(v)$ into $F_Y(v)$ is further substituted into (28) to arrive at the CDF in Theorem 6. By substituting $p_C(v)$ and $F_C(v)$ from Appendix A into $p_R(v)$, we obtain the PDF.

APPENDIX D
ON EXTREME VALUE THEORY

In this appendix, we review a few important results from the extreme value theory that are relevant to this paper. Extreme value theory is a branch of statistics. It primarily deals with the problem of finding the probability distributions of the order statistics (maxima or minima) of $n$ i.i.d. random variables when $n$ tends to a sufficiently large value. This core problem has been well-documented in [42][45][51-53].

Let $\{Z_i\}_{i=1}^n$ be a sequence of i.i.d. random variables and $X_n := \min(Z_1, Z_2, \cdots, Z_n)$ and $Y_n := \max(Z_1, Z_2, \cdots, Z_n)$. The Fisher-Tippett theorem [39, p. 121] says that the limit (asymptotic) CDFs of $X_n$ or $Y_n$ belong to one of the following extreme value CDFs: Gumbel, Weibull or Fréchet. It is worth to mention here that the CDF for the maxima and the minima may not always converge to the same type of distribution. For example, the CDF of the minimum of i.i.d. Chi-square random variables converges (in distribution) to the Weibull, whereas the maximum converges to the Gumbel [45, Table 9.5].

*1. Minimum of Chi-square random variables*

Let $\{\chi_i^2\}_{i=1}^n$ be a sequence of i.i.d. Chi-square random variables with $M$ degrees of freedom and let $X_n = \min_{i=1,2,\cdots,n} \chi_i^2$. It is known [45, Table 9.5] that the CDF of $X_n$ converges to the scaled and translated Weibull CDF:

$$F_X^\infty(u) \to W_L\left(\frac{u-p_n}{q_n}\right) = \begin{cases} 1 - \exp\left(-\left(\frac{u-p_n}{q_n}\right)^\beta\right) & u \geq p_n \\ 0 & u < p_n \end{cases}$$
(30)

where $p_n$, $q_n$ and $\beta$ are the parameters of the distribution. We now state the asymptotic distribution for the left RIV.

***Lemma 14 Asymptotic distribution for the left RIV:*** Consider the left RIV $\Delta_K^L := 1 - \min_{i=1,2,\cdots,n} \chi_i^2$ with $n = \binom{N}{K}$. With the scaling constant $q_n = \frac{6}{M} \exp\left(\frac{2}{M}\left[\ln \Gamma\left(\frac{M}{2}\right) + \ln \frac{M}{2} - \ln\binom{N}{K}\right]\right)$, the location constant $p_n = 0$ and with the shape constant $\beta = \frac{M}{2}$, the CDF of the left RIV converges to $1 - W_L\left(\frac{1-u-p_n}{q_n}\right)$.

Proof: Since the CDF $F_X^\infty(u)$ of $X_n$ converges to $W_L\left(\frac{u-p_n}{q_n}\right)$ (see (30)), the CDF of the left RIV $F_L^\infty(u)$ converges to $1 - F_X^\infty(1-u)$. By substituting $F_X^\infty(u)$, we obtain the asymptotic distribution of the left RIV as

$$F_L^\infty(u) = 1 - W_L\left(\frac{1-u-p_n}{q_n}\right)$$
$$= \exp\left(-\left(\frac{1-u}{q_n}\right)^\beta\right).$$

We need to find the constants of the CDF. They can be derived by using the CDF $F_C(x)$ in Appendix A.

Location constant: The location parameter for the Weibull distribution is $p_n = v(F) = \inf\{x \mid F_C(x) > 0\}$, which is the lower end of the CDF. Since the Chi-square distribution is supported on $[0, \infty)$, $p_n = v(F) = 0, \forall n$.

Scale constant: The scale parameter for the Weibull is $q_n = F_C^{-1}\left(\frac{1}{n}\right) - v(F)$. Since $v(F) = 0$, $q_n = F_C^{-1}\left(\frac{1}{n}\right)$. To find $F_C^{-1}\left(\frac{1}{n}\right)$, we write the CDF $F_C(x)$ in Taylor series as

$$F_C(x) = \frac{\gamma\left(\frac{M}{2}, \frac{M}{2}x\right)}{\Gamma\left(\frac{M}{2}\right)} = \frac{1}{\Gamma\left(\frac{M}{2}\right)} \sum_{i=0}^\infty \frac{(-1)^i \left(\frac{M}{2}x\right)^{i+\frac{M}{2}}}{i!\left(i+\frac{M}{2}\right)}.$$

We need to find $x$ such that $F_C(x) = \frac{1}{n}$. Since $\frac{1}{n}$ approaches a very small value as $n \to \infty$, the value of $x$ must be very small as well. So, we approximate $F_C(x)$ by using only the first term of the series. That is, $F_C(q_n) = \frac{1}{n} = \frac{\left(\frac{M}{2}q_n\right)^{\frac{M}{2}}}{\frac{M}{2}\Gamma\left(\frac{M}{2}\right)}$ from which we determine $q_n = \frac{2}{M}\left[\frac{\frac{M}{2}\Gamma\left(\frac{M}{2}\right)}{n}\right]^{\frac{2}{M}}$. We observe via the Newton's iterative method that if we include more number of terms in order to approximate $F_C(x)$, the value of $q_n$ (with one term approximation) scales up to a factor of 3. Hence, we multiply the $q_n$ by 3. By substituting $n = \binom{N}{K}$ and expressing $q_n$ in terms of logarithms we obtain the $q_n$ mentioned in Theorem 7.

Shape parameter: We obtain the shape parameter for the Weibull by evaluating the following limit [45]:

$$\lim_{z \to -\infty} \frac{F\left(v(F) - \frac{1}{zx}\right)}{F\left(v(F) - \frac{1}{z}\right)} = x^{-\beta}, \quad \beta > 0.$$

We evaluate the above limit by considering the CDF $F_C(x)$. Since $v(F_C) = 0$ and hence,

$$\lim_{z \to -\infty} \frac{F_C\left(-\frac{1}{zx}\right)}{F_C\left(-\frac{1}{z}\right)} = \lim_{z \to -\infty} \frac{\frac{1}{\Gamma\left(\frac{M}{2}\right)} \sum_{i=0}^{\infty} \frac{(-1)^i \left(-\frac{M}{2zx}\right)^{i+\frac{M}{2}}}{i!\left(i+\frac{M}{2}\right)}}{\frac{1}{\Gamma\left(\frac{M}{2}\right)} \sum_{i=0}^{\infty} \frac{(-1)^i \left(-\frac{M}{2z}\right)^{i+\frac{M}{2}}}{i!\left(i+\frac{M}{2}\right)}} \quad (31)$$

$$= x^{-\frac{M}{2}}$$

From (31) we judge the shape constant as $\beta = \frac{M}{2}$. ∎

*2. Support of the PDF of the left RIV*

Here we prove Lemma 8. We denote the LESP as $u_{\text{left}}$ and and the UESP as $u_{\text{right}}$, which are defined as follows:

$$u_{\text{left}} := \inf\{u \in [0,1]: F_L^{\infty}(u) \geq \varepsilon\}$$
$$u_{\text{right}} := \inf\{u \in [0,1]: F_L^{\infty}(u) \geq 1-\varepsilon\}$$

Then, the support of the PDF is $S_L = \{u : u_{\text{left}} \leq u \leq u_{\text{right}}\}$. Since $F_L^{\infty}(u)$ is known, we can obtain $u_{\text{left}}$ by setting $F_L^{\infty}(u) = \varepsilon$ and solving for $u$. This yields $u_{\text{left}} = 1 - q\left[\ln \frac{1}{\varepsilon}\right]^{\frac{1}{\beta}}$. Similarly, by setting $F_L^{\infty}(u) = 1-\varepsilon$ and solving for $u$ gives $u_{\text{right}} = 1 - q\left[\ln \frac{1}{1-\varepsilon}\right]^{\frac{1}{\beta}}$.

*3. Maximum of Chi-square random variables*

Let $\{\chi_i^2\}_{i=1}^n$ be a sequence of i.i.d. Chi-square random variables with $M$ degrees of freedom with the CDF $F_C(x)$ and let $Y_n = \max_{i=1,2,\cdots,n} \chi_i^2$. It is known [42, p. 156] that the CDF of $Y_n$ converges to the scaled and translated Gumbel CDF:

$$F_Y^{\infty}(v) \to G_R\left(\frac{v-l_n}{s_n}\right) = \exp\left(-\exp\left(-\frac{v-l_n}{s_n}\right)\right), \quad v \in \mathbb{R} \quad (32)$$

where $s_n$ and $l_n$ are the constants of the Gumbel CDF. We now state the asymptotic CDF for the right RIV.

***Lemma 15 Asymptotic distribution for the right RIV:*** Consider the right RIV $\Delta_K^R := \max_{i=1,2,\cdots,n} \chi_i^2 - 1$ with $n = \binom{N}{K}$. With the scaling constant $s_n = 2\sigma^2$ and the location constant $l_n = s_n\left(\ln n + \left(\frac{M}{2}-1\right)\ln \ln n - \ln \Gamma\left(\frac{M}{2}\right)\right)$, the CDF of the right RIV converges in distribution to $G_R\left(\frac{v+1-l_n}{s_n}\right)$.

Proof: Since the CDF $F_Y^{\infty}(v)$ of $Y_n$ converges to $G_R\left(\frac{v-l_n}{s_n}\right)$, the CDF of the right RIV converges to $F_R^{\infty}(v) = F_Y^{\infty}(v+1)$. Hence, by using (32) we obtain

$$F_R^{\infty}(v) = F_Y^{\infty}(v+1)$$
$$= \exp\left(-\exp\left(-\left(\frac{v+1-l}{s}\right)\right)\right)$$

The derivation of the scaling and the location constants can be referred from [52, p. 72] [42, Table 3.4.4]. ∎

APPENDIX E
SUFFICIENCY OF I.I.D REPRESENTATION FOR RIC PREDICTION

In this Appendix, we aim to prove that the i.i.d. representation of the RICs ((5) and (6)) is sufficient for both the RIC predicted region and the strong equivalence region. The representation of the RIVs in (5) and (6) when any two ratio random variables $R(\mathcal{S}_i)$ and $R(\mathcal{S}_j), i \neq j$ are not independent in general. The use of i.i.d. representation of the RICs done in this paper for the purpose of RIC prediction, therefore, needs careful justification.

<u>Sufficiency of the i.i.d. representation for the RIC predicted region.</u> The notations for the critical functions of the RICs (left and right) for i.i.d. and the dependent representations are given in Table 2.

Table 2: Notations for the Critical functions for i.i.d and dependent representation

| Representation | Notations for Critical Functions | |
|---|---|---|
| | Left RIC | Right RIC |
| i.i.d | $u^I(\rho,\delta)$ | $v^I(\rho,\delta)$ |
| Dependent | $u^C(\rho,\delta)$ | $v^C(\rho,\delta)$ |

We first deal with the left RIV. Let $F_L^C(x)$ be the CDF of the left RIV using the i.i.d. representation and $F_L^I(x)$ be the CDF using the dependent representation. We prove that $F_L^I(x) \leq F_L^C(x), \forall x$. This implies that the critical functions $u^I(\rho,\delta)$ and $u^C(\rho,\delta)$ obey $u^I(\rho,\delta) \leq u^C(\rho,\delta), \forall \rho, \delta$. Then, the effect of this fact on the level curves $\{(\rho_I,\delta): u^I(\rho,\delta) = c\}$ and $\{(\rho_C,\delta): u^C(\rho,\delta) = c\}$ is that $\rho_I \leq \rho_C, \forall \delta$. That is, the level curve of the i.i.d. representation is below that of the dependent representation. Thus, the i.i.d. representation of the left RIC is sufficient for the RIC predicted region.

It remains to prove $F_L^I(x) \leq F_L^C(x), \forall x$ or equivalently $\Pr\{1 - \min R^I \leq x\} \leq \Pr\{1 - \min R^C \leq x\}$, where $R^C = R_{\mathcal{S}_1}^C, R_{\mathcal{S}_2}^C, \cdots, R_{\mathcal{S}_P}^C$ is the collection of $P = \binom{N}{K}$ dependent random variables and $R^I = R_{\mathcal{S}_1}^I, R_{\mathcal{S}_2}^I, \cdots, R_{\mathcal{S}_P}^I$ is the collection of $P$ i.i.d. random variables.

Our proof is given in two steps. In the first step, we prove $\Pr\{\min R_S^I \leq x\} \geq \Pr\{\min R_S^C \leq x\}, \forall x,$ and in the second step we prove $\Pr\{1 - \min R_S^I \leq x\} \leq \Pr\{1 - \min R_S^C \leq x\}, \forall x.$

**Step 1.**

We start with R.H.S. $\Pr\{\min R_S^C \leq x\}$. By the definition of minimum of random variables, we can write the probability as

$$\Pr\{\min R_S^C \leq x\} = 1 - \Pr\{\min R_S^C > x\}$$
$$= 1 - \Pr\{R_{S_1}^C > x, R_{S_2}^C > x, \cdots, R_{S_P}^C > x\}$$

Now the probability $\Pr\{R_{S_1}^C > x, R_{S_2}^C > x, \cdots, R_{S_P}^C > x\}$ can be written in terms of indicator function [] as

$$\Pr\{R_{S_1}^C > x, R_{S_2}^C > x, \cdots, R_{S_P}^C > x\} = \mathbb{E}\left[\mathbf{1}_{\{R_{S_1}^C > x, R_{S_2}^C > x, \cdots, R_{S_P}^C > x\}}\right]$$

For the case of $P = 2$,

$$\Pr\{R_{S_1}^C > x, R_{S_2}^C > x\} = \mathbb{E}\left[\mathbf{1}_{\{R_{S_1}^C > x, R_{S_2}^C > x\}}\right]$$
$$= \mathbb{E}\left[\mathbf{1}_{\{R_{S_1}^C > x\}} \mathbf{1}_{\{R_{S_2}^C > x\}}\right]$$
$$= \prod_{k=1}^{2} \mathbb{E}\left[\mathbf{1}_{\{R_{S_k}^C > x\}}\right] + \mathbb{C}ov\left(\mathbf{1}_{\{R_{S_1}^C > x\}}, \mathbf{1}_{\{R_{S_2}^C > x\}}\right)$$
$$= \prod_{k=1}^{2} \mathbb{E}\left[\mathbf{1}_{\{R_{S_k}^I > x\}}\right] + \mathbb{C}ov\left(\mathbf{1}_{\{R_{S_1}^C > x\}}, \mathbf{1}_{\{R_{S_2}^C > x\}}\right)$$
$$\geq \prod_{k=1}^{P} \mathbb{E}\left[\mathbf{1}_{\{R_{S_k}^I > x\}}\right]$$
$$= \Pr\{R_{S_1}^I > x, R_{S_2}^I > x\}$$

The inequality follows from the fact that the covariance of two dependent Chi-square random variables (square of bi-variate Gaussian) is positive. Similarly, the three variable case $P = 3$ can be written using the two random variable case as

$$\Pr\{R_{S_1}^C > x, R_{S_2}^C > x, R_{S_3}^C > x\} = \mathbb{E}\left[\mathbf{1}_{\{R_{S_1}^C > x, R_{S_2}^C > x\}} \mathbf{1}_{\{R_{S_3}^C > x\}}\right]$$
$$= \mathbb{E}\left[\mathbf{1}_{\{R_{S_1}^C > x, R_{S_2}^C > x\}}\right] \mathbb{E}\left[\mathbf{1}_{\{R_{S_3}^C > x\}}\right]$$
$$+ \mathbb{C}ov\left(\mathbf{1}_{\{R_{S_1}^C > x, R_{S_2}^C > x\}}, \mathbf{1}_{\{R_{S_3}^C > x\}}\right)$$

Now, expanding the first expectation using the results from $P = 2$, we obtain

$$\Pr\{R_{S_1}^C > x, R_{S_2}^C > x, R_{S_3}^C > x\} = \mathbb{E}\left[\mathbf{1}_{\{R_{S_1}^C > x\}}\right] \mathbb{E}\left[\mathbf{1}_{\{R_{S_2}^C > x\}}\right]$$
$$\mathbb{E}\left[\mathbf{1}_{\{R_{S_3}^C > x\}}\right]$$
$$+ \mathbb{C}ov\left(\mathbf{1}_{\{R_{S_1}^C > x\}}, \mathbf{1}_{\{R_{S_2}^C > x\}}\right) \mathbb{E}\left[\mathbf{1}_{\{R_{S_3}^C > x\}}\right]$$
$$+ \mathbb{C}ov\left(\mathbf{1}_{\{R_{S_1}^C > x, R_{S_2}^C > x\}}, \mathbf{1}_{\{R_{S_3}^C > x\}}\right)$$
$$\geq \prod_{k=1}^{3} \mathbb{E}\left[\mathbf{1}_{\{R_{S_k}^I > x\}}\right]$$
$$\geq \Pr\{R_{S_1}^I > x, R_{S_2}^I > x, R_{S_3}^I > x\}$$

Repeating the process for $P$ random variables case, we obtain

$$\Pr\{R_{S_1}^C > x, R_{S_2}^C > x, \cdots, R_{S_P}^C > x\} \geq \prod_{k=1}^{P} \mathbb{E}\left[\mathbf{1}_{\{R_{S_k}^I > x\}}\right]$$
$$= \Pr\{R_{S_1}^I > x, R_{S_2}^I > x, \cdots, R_{S_P}^I > x\}$$

Hence,

$$\Pr\{\min R^C > x\} = \Pr\{R_{S_1}^C > x, R_{S_2}^C > x, \cdots, R_{S_P}^C > x\}$$
$$\geq \Pr\{R_{S_1}^I > x, R_{S_2}^I > x, \cdots, R_{S_P}^I > x\}$$
$$= \Pr\{\min R^I > x\}$$

And therefore,

$$\Pr\{\min R^C \leq x\} \leq \Pr\{\min R^I \leq x\}, \forall x.$$

**Step 2.**

Now, let us consider, $\Pr\{1 - \min R_S^I \leq x\}$

$$\Pr\{1 - \min R_S^I \leq x\} = \Pr\{\min R_S^I \geq 1 - x\}$$
$$= 1 - \Pr\{\min R_S^I \leq 1 - x\}$$
$$\leq 1 - \Pr\{\min R_S^C \leq 1 - x\}$$
$$= \Pr\{1 - \min R_S^C \leq x\}$$

Thus, we note that $\Pr\{1 - \min R_S^I \leq x\} \leq \Pr\{1 - \min R_S^C \leq x\}$. This implies that the CDFs of the left RIV satisfy the following inequality $F^I(x) \leq F^C(x), \forall x$ which is what we aimed at proving.

For the right RIV, we need to prove $\Pr\{\max R^I - 1 \leq y\} \leq \Pr\{\max R^C - 1 \leq y\}$. For this one, we first prove $\Pr\{\max R_S^C \leq y\} \geq \Pr\{\max R_S^I \leq y\}, \forall y$ by following the same steps for the left RIV and then prove $\Pr\{\max R^I - 1 \leq y\} \leq \Pr\{\max R^C - 1 \leq y\}$.

In summary, we proved that $F_L^I(x) \leq F_L^C(x), \forall x$ and $F_R^I(x) \leq F_R^C(x), \forall x$. Thus, we have shown that the i.i.d. representation is sufficient for obtaining the RIC predicted region.

Sufficiency of i.i.d. representation for the strong equivalence region: Consider the critical functions for the left RIC and the right RIC using the i.i.d. and dependent representation in Table I. The FL function for these representation are given, respectively, by

$$\mu^I(\rho, \delta) := \frac{1 + \sqrt{2}}{4}\left(\frac{1 + v^I(2\rho, \delta)}{1 - u^I(2\rho, \delta)} - 1\right) \text{ and }$$

$$\mu^C(\rho, \delta) := \frac{1 + \sqrt{2}}{4}\left(\frac{1 + v^C(2\rho, \delta)}{1 - u^C(2\rho, \delta)} - 1\right)$$

For brevity, we drop the arguments of all the functions in the subsequent discussions. If $\mu^I \leq \mu^C, \forall \rho, \delta$ then FL function $\mu^I$ is sufficient to predict the strong equivalence region. That is, if $\mu^I \leq \mu^C, \forall \rho, \delta$ then the level curve $\{(\rho_I, \delta): \mu^I(\rho, \delta) = c\}$ is always below the level curve $\{(\rho_C, \delta): \mu^C(\rho, \delta) = c\}, \forall c$. Thus, it is sufficient to prove $\mu^I \leq \mu^C, \forall \rho, \delta$. We already know that $u^I \leq u^C, \forall \rho, \delta$ and $v^I \leq v^C, \forall \rho, \delta$. In order to prove $\mu^I \leq \mu^C, \forall \rho, \delta$ it is sufficient to prove that the ratio $\frac{\mu^I}{\mu^C} \leq 1 \ \forall \rho, \delta$.

$$\frac{\mu^I}{\mu^C} = \frac{\left(\frac{1+v^I}{1-u^I} - 1\right)}{\left(\frac{1+v^C}{1-u^C} - 1\right)}$$

$$= \frac{(v^I + u^I)(1 - u^C)}{(v^C + u^C)(1 - u^I)}$$

$$\leq \frac{(v^I + u^I)(1 - u^C)}{(v^C + u^C)(1 - u^C)} \quad (\text{since } u^I \leq u^C)$$

$$\leq 1 \quad (\text{since } v^I + u^I \leq v^C + u^C)$$

Hence, $\mu^I(\rho, \delta) \leq \mu^C(\rho, \delta), \forall \rho, \delta$ is proved and thus, i.i.d. representation is sufficient for strong equivalence region as well.